# Emergence and Computation at the Edge of Classical and Quantum Systems

*Ignazio Licata*
*Isem, Institute for Scientific Methodology, Palermo, Italy*
Ignazio.licata@ejtp.info

**Abstract** : The problem of emergence in physical theories makes necessary to build a general theory of the relationships between the observed system and the observing system. It can be shown that there exists a correspondence between classical systems and computational dynamics according to the Shannon-Turing model. A classical system is an informational closed system with respect to the observer; this characterizes the emergent processes in classical physics as phenomenological emergence. In quantum systems, the analysis based on the computation theory fails. It is here shown that a quantum system is an  informational open system with respect to the observer and able to exhibit processes of observational, radical emergence. Finally, we take into consideration the role of computation in describing the physical world.

1. **Introduction**

The study of the complex behaviors in systems is one of the central problems in Theoretical Physics. Being related to the peculiarities of the system under examination, the notion of complexity is not univocal and largely interdisciplinary, and this accounts for the great deal of possible approaches. But there is a deeper epistemological reason which justifies such intricate "archipelago of complexity": the importance of the observer's role in detecting complexity, that is to say those situations where the system's collective behaviors give birth to structural modifications and hierarchical arrangements. This consideration directly leads to the core of the emergence question in Physics.
  We generally speak of emergence when we observe a "gap" between the formal model of a system and its behaviors. In other words, the detecting of emergence expresses the necessity or, at least, the utility for the creation of a new model able to seize the new observational ranges.So the problem of the relationship among different description levels is put and two possibile situations arise: 1) phenomenological emergence, where the observer operates a "semantic" intervention according to the system's new behaviors, and aiming at creating a new model –  choosing the state variables and dynamical description – which makes the description of the observed processes more convenient. In this case the two descriptive levels can be always – in principle – connected by opportune "bridge laws", which carry out such task by means of a finite quantity of syntactic information; 2) radical emergence, where the new description cannot be connected to the initial model. Here we usually observe a breaking of the causal chain (commonly describable through opportune symmetries), and irreducible forms of unpredictability. Hence, the link between the theoretical corpus and the new model could require a different kind of semantics of the theory, such as a new interpretation and a new arrangement of the basic propositions and their relationships.
  Such two distinctions have to be considered as a mere exemplification, actually more varied and subtler intermediate cases can occur. The relationships between Newtonian Dynamics and the concept of entropy can be taken into consideration as an example of phenomenological emergence. The laws of Classical Dynamics are time-reversal, whereas entropy defines a "time arrow". In order to connect the two levels, we need a new model based on Maxwell-Boltzmann statistics as well as on a refined probabilistic hypothesis centered, in turn, on space-time symmetries - because of the space-time isotropy and homogeneity there do not exist points, directions or privileged instants in a de-correlation process between energetic levels-. So, a "conceptual bridge" can be built between the particle description and entropy, and consequently between the microscopic and macroscopic analysis of the system. But this connection does not cover all the facets of the problem, and thus we cannot regard it as a "reduction" at all. In fact in some cases, even within the closed formulation of classical physics, entropy can decrease locally, and after all the idea to describe a perfect gas in molecular terms would never cross anybody's mind!

Another example regards the EPR-Bell correlations and the non-locality role in Quantum Mechanics. Within the Copenhagen Interpretation the non-local correlations are experimentally observed but they are not considered as facts of the theory. In the Bohm Interpretation the introduction of the quantum potential makes possible to bring non-locality within the theory. It should not be forgotten that historically the EPR question comes out as a *gedankenexperiment* between Einstein and Bohr about the "elements of physical reality" in Quantum Mechanics. Only later, thanks to Bohm analysis and Bell's Inequality on the limits of local hidden variable theories, such question developed into an experimental matter. Nor Einstein neither Bohr would expect to observe really the "ghost-like-action-at-a-distance". It is useful to remember that in Bohm theory the introduction of non-locality does not require any additional formal hypotheses but the standard apparatus provided by the Schrödinger equation. Besides, if on one hand the new interpretative perspective provides a different comprehension of the theory, on the other hand it puts some problems about the so-called "pacific coexistence" between Restricted Relativity and Quantum Mechanics.

In both of the briefly above-examined cases we can see how the phenomenological and radical features of emergence are strongly intertwined with the development dynamics of physical theories and how the general problem of emergence points up questions of fundamental importance for the physical world description, such as the updating mechanism of the theories and the crucial role of the observer in choosing models and their interpretations. In particular, it is worth noticing that the relationship between the observer and the observed is never a merely "one-way" relationship and it is unfit to be solved in a single direction, which would lead to epistemological impoverishment. This relationship has rather to be considered as an adaptive process in which the system's internal logic meets our modalities to acquire information about it in order to build theories and interpretations able to shape up a system's description.

The problems related to the emergence theory, conceived as a general theory of the relationships between observing system and observed system, will be here taken into consideration, and will be tested on some evolution models of both classical and quantum systems. Finally, we will develop some considerations about the logic limits of the theories and the computability role in describing the physical world.

## 2. The Observers in Classical Physics: Continuous Systems

For our aims, an informational or logical closed formal system will be intended as a model of physical system such that: 1) the state variables and the evolution laws are individuated; 2) it is always possible to obtain the values of the variable states at each instant; 3) thanks to the information obtained by the abovementioned two points, it is always possible to connect univocally the input and the output of the system and to forecast its asymptotic state. So, *a logical closed formal system is a deterministic system* with respect to a given choice of the state variables.

Let us consider a classical system and see how we can regard it as logical closed with respect to the observation procedures and its ability to show emergent processes. To be more precise, the values of the state variables express the *intrinsic* characteristic properties of the classical object and they are not affected by the measurement. Such fact, as it is known, can be expressed by saying that in a classical system the measurement made on all state variables are *commutative* and *contextually compatible*, i.e. all the measurement apparatuses connected to different variables can always be used without interfering one with the other and without any loss of reciprocal information. This assumption, supported by the macroscopic observations, leads to the idea of a biunivocal correspondence between the system, its states and the outcomes of the measurements. Hence, the logic of Classical Physics is Boolean and orthocomplemented, and it formalizes the possibility to acquire complete information about any system's state for any time interval. The description of any variation in the values of the state variables, at each space-time interval, defines the *local evolutionary* feature of a classical system, either when it is "embedded" within the structure of a system of differential equations or within discrete transition rules.

The peculiar independence of a classical system's properties from the observer has deep consequences for the formal structure of classical physics. It is such independence which characterizes the system's *local, causal determinism* as well as the principle of *distinguishability of states* in the phase space according to Maxwell-Boltzmann distribution function. This all puts very strong constraints to the informational features of classical physics and its possibility to show emergence.

The correspondence between the volume in a classical system's phase space and Shannon information via Shaw's Theorem (Shaw, 1981) allows to combine the classification of the thermodynamic schemes (isolated, closed, open) in a broader and more elaborate vision. Three cases are possible:

a) *Information-conserving* systems (Liouville's Theorem);

b) *Information-compressing* systems, ruled by the second principle of thermodynamics and consequently by the *microscopic principle of correlation weakening* among the constituents of the system individuated by the Maxwell-Boltzmann statistics (see Rumer & Ryvkin, 1980). These systems, corresponding to the closed ones, have a finite number of possible equilibrium states. By admitting a more general conservation principle of information and suitably redefining the system's boundaries for the (b)-type systems, we can connect the two kind of systems (a) and (b), and so coming to the conclusion that in the latter a passage from macroscopic information to microscopic information takes place;

c) *Information-amplifying* systems which show definitely more complex behaviors. They are non-linear systems where the variation of a given order parameter can cause macroscopic structural modifications. In these systems, the time dependence between the V volume of the phase space and the I information is given by: $dI/dt = (1/V)dV/dt$. The velocity of information production is strictly linked to the kind of non-linearity into play and can thus be considered as a measure of complexity of the systems. Two principal classes can be individuated: c-1) Information-amplifying systems in polynomial time, to which the dissipative systems able to show self organization processes belong (Prigogine, 1994; Haken, 2004) ; c-2) Information-amplifying systems in exponential time; they are structurally unstable systems (Smale, 1966), such as the deterministic chaotic systems. Both the information-amplifying types belong to the open system classes, where an infinite number of possible equilibrium states are possible.

Despite their behavioral diversity, the three classical dynamic systems formally belong to the class of logical closed models, i.e., because of the deep relationships between local determinism, predictability and computation, they allow to describe the system by means of recursive functions. If we consider the evolution equations as a local and intrinsic computation,we will see that in all of the three examined cases it is possible to characterize the incoming and outgoing information so as to define univocally the output/input relationships at each time (Cruchtfield, 1994). In dissipative systems, for example, information about the self-organized stationary state *is already contained* in the structure of the system's equations. The actual setting up of the new state is due to the variation of the order parameter in addition to the boundary conditions. Contrary to what is often stated, even the behavior of structural unstable systems – and highly sensible to initial conditions- is asymptotically predictable as strange attractor. What is lacking is the connection between global and local predictability, but we can always follow step-by-step computationally the increase of information within a predefined configuration. Our only limit is the power of resolution of the observation tool and/or the number of computational steps. In both cases we cannot speak of intrinsic emergence, but of emergence as *detection of patterns*.

### 3. The Observers in Classical Physics: Discrete Systems

The discrete systems such as Cellular Automata (CA) (Wolfram, 2002) represent interesting cases. They can be considered as classical systems as well, because the information on the evolution of the system's states is always available for the observer in the same way we saw for the continuous systems. On the other hand, their features are quite different than those of such systems in relation to emergent behavior.

The Wolfram-Langton classification (Langton, 1990) identifies four fundamental classes of cellular automata. At the $\lambda$ parameter's varying- a sort of generalized energy – they show up the following sequence:

Class I (evolves to a homogeneous state)→ Class II (evolves to simple periodic or quasi-periodic patterns)→Class IV (yields complex patterns of localized structures with a very long transient, for ex. Conway *Life Game*)→Class III (yields chaotic aperiodic patterns).

It is known that cellular automata can realize a Universal Turing Machine (UTM). To this general consideration, the Wolfram-Langton classification adds the analysis of the evolutionary behaviors of discrete systems, so building an extreme interesting bridge between the theory of dynamical systems and its computational facets.

The I, II, III classes can be directly related to the information-compressing systems, the dissipative-like polynomial amplifiers and the structural unstable amplifiers, respectively.This makes the CA a powerful tool in simulating physical systems and the privileged one among the discrete models. The correspondence between a continuous system and the class IV appears to be more problematic. This class looks rather like an intermediate dynamic typology between unstable systems and dissipative systems, able to show a strong peculiar order/chaos mixture. It suggests they are systems which exhibit emergence *on the edge of chaos* in special way with respect to the case of the continuous ones. (Bak et al., 1988).

Although this problem is still questionable from the conceptual and formal viewpoint, it is possible to individuate at least a big and significant difference between CA and continuous systems. We have seen that information is not erased in information-compressing systems, but a passage from macroscopic to microscopic information takes place. Therefore, the not time reversal aspects of the system belong more to the phenomenological emergence of entropy at descriptive level than the local loss of information about its states. In other words, for the conservation principle and the distinguishability of states, the information about any classical particles is always available for observation. It is not true for CA, there the irreversible erasing of local states can take place, such as in some interactions among *gliders* in *Life*. The situation is analogous to the middle-game in chess, where some pieces have been eliminated from the game. In this case, it is impossible to univocally reconstruct the opening initial conditions. Nevertheless, it is always possible to individuate *at least one computational path* able to connect the initial state to the final one, and it is possible to show, thanks to the *finite* number of possible paths, that one of such paths must be the one that the system has actually followed. Consequently, if on the one hand the erasing of information in CA suggests more interesting possibilities of the discrete emergence in relation to the continuous one, on the other its characteristics are not so marked to question about the essential classical features of the system. In fact, in more strictly physical terms, it is also possible for the observer to locally detect the *state erasing* without losing the global describability of the dynamic process in its causal features.

Some interesting formal analogies between the unpredictability of structural unstable systems and the *halting problem* in computation theory can be drawn. In both cases, there is no correlation between local and global predictability, and yet the causal determinism linked to the observer's possibility to follow step-by-step the system's evolution is never lost. Far from simply being the base for a mere simulation of classical systems, such point illuminates the deep connection between computation and classical systems. Our analysis has provided broad motives for justifying the following definition of classical system: *a classical system is a system whose evolution can be described as an intrinsic computation in Shannon-Turing sense*. It means that any aspect of the system's "unpredictability" is not connected to the causal structure failing and any loss of information can be individuated locally by the observer. All this is directly linked to what we called the classical object's *principle of indifference* to the measurement process and can be expressed by saying that *classical systems are informational closed with respect to the observer.*

Such analysis, see in the following, is not valid for the quantum systems. This is one of the reasons which makes the introduction of the indeterminism in discrete systems quite problematic (Friedkin, 1992; Licata, 2001).

## 4. Quantum Events and Measurement Processes

*4.1) The Centrality of Indeterminism*

The Heisenberg Uncertainty Principle is the essential conceptual nucleus of Quantum Physics and characterizes both its entire formal development and the debate on its interpretations.

For a quantum system, the set of values of the state variables is not accessible to the observer at each moment. It is so necessary to substitute the "state variable" concept with the system state notion, characterized by the wave-function or state vector $|\Psi\rangle$. The relation between the system state and the variables is given by the Born rule. For example, given a state variable q we can calculate its probability value through the state vector's general form. It leads to explicitly introduce in the formalism the measurement operations as operators which are - differently from the classical ones – *non-commutative* and *contextually incompatible*. So, Quantum Logic differs from the Classical one because it is relatively orthocomplemented and not Boolean, thus orthomodular (D'Espagnat,1999). A direct formal consequence is that for the quantum objects there exists a *Principle of indistinguishability of states* which characterizes the Fermi-Dirac as well as the Bose-Einstein quantum statistics.

In this way, the observer takes on a radical new role, which can be clarified by means of the relational notion of *quantum event* (Healy, 1989; Bene, 1992; Rovelli, 1996; for the notion of quantum contextuality as a *frame* see also Griffiths, 1995; for the collapse problem from the logic viewpoint see: Dalla Chiara, 1977).

Let us consider a quantum system S and an observer O. The interaction between the two systems, here indicated as O+S, will give a certain value *q* of the corresponding state variable q. So, we can say – by paraphrasing Einstein – that the quantum event *q* is the *element of physical reality* of the system S with respect to O. If we describe S and O as two quantum systems, for ex. two-state systems - another system O' will find the value *q'* relative to the system S+O. Therefore, a quantum event always expresses an interaction between two systems and, according to the relational approach, this is all we can know about the physical world, because a hypothetical comparison between O and O' *will be a quantum event, too*.

Despite the fact that it cannot be fully considered as an "interpretation" yet, the relational view clearly shows the irreducibility of the role of the observer in QM as well as the new symmetry with the observed system. In spite of such radically not classic feature, the so-called "objective reality" described by the quantum world has not to be questioned at all; it is precisely the peculiar nature of the quantum objects which imposes a new relationship with measurements on the theory and leads to a far different notion of event than the classical one.

*4.2) The Collapse Postulate*

When we consider the "collapse postulate" of the wave function, a radical asymmetry will be unavoidable. In this case, we have to consider on the one hand the temporal evolution of the wave function **U**, provided by the rigorously causal, deterministic and time-reversal Schrödinger equation, and on the other the reduction processes of the state vector **R**, where the measurement's outcome is macroscopically fixed, such as the position of an electron on a photographic plate. The processes **R** are not-causal and time asymmetrical.

Different standpoints are possible about the role of the processes **R** in QM, it recalls the 1800s mathematical debate about the structure of Euclidean geometry and in particular the fifth postulate position. Without any claim to exhaustiveness, we can individuate here three main standpoints about **R**:

A. The wave function contains the available information on the physical world in probabilistic form; the wave function is not referred to an "objective reality", but - due to the intrinsically relational features of the theory - only to what we can say about reality. Consequently, the "collapse postulate" is simply an expression of our peculiar knowledge of the world of quantum objects;
B. The wave function describes what actually happens in the physical world and its probabilistic nature derives from our perspective of observers; **R**, as well as the entire QM, is the consequence of the fact that the most part of the needed knowledge is structurally unavailable;
C. The wave function partially describes what happens in the physical processes; in order to comprehend its probabilistic nature and the postulate **R** in particular, we need a theory connecting **U** and **R**.

The above three standpoints are subtly different and yet connected. To clarify these interrelations we will give some examples. (A) reflects the traditional Copenhagen interpretation. We are not interested here in the old ontological debate about reality, but rather we are going to focus on the formal-logical structure of each theory. According to the supporters of (A) – or at least one of its manifold forms – QM is coherent and complete, and the collapse postulate assures the connection between **U** and the observations, even if this connection cannot be deduced by **U**. It means that in order to provide a quantum event, the "Von Neumann chain" breaks at the observer level. The price to pay for such readings of the theory is that the non-local correlations come out as a compromise between causality (conservation laws and Relativity) and quantum casuality ( **R** irreducibility).

In the class (B) quite different interpretations are taken into consideration, such as the Everett Theory on the relative states (DeWitt & Graham, 1973; Deutsch, 1985) and the Bohm-Hiley theory of Implicate Order (Bohm & Hiley, 1995). The Everett Many Worlds Theory regards **U** as the description of the splitting of the branches in the multiverse (Deutsch, 1998), and **R** simply as the observer's act of classically detecting its own belonging to one of the branches of the quantum multiverse by the measurement. The Everett Theory can be considered to lie at the bottom of the current relational approaches as well as, indirectly, the so-called decoherent histories and, in general, many of the quantum cosmology conceptions (Gell-Mann & Hartle, 1993; Omnès, 1994). Unfortunately, the idea of the multiverse does not solve all the **R**-related problems.The real problem is to clarify why the universes interfere one with the other and show entanglement with respect to any observer at a given scale. So the idea to reduce the quantum superposition to a classical collection of observers fails. That is why, in the decoherent histories, the problem shifts to the emergence– a coarse-grained one, at least – of classicality from quantum processes.

For our aims it is particularly relevant the recent observation that a coherent theory of multiverse implies the existence of objects carrying not classical information, such as the *ghost-spinors* (Palesheva, 2001; 2002; Guts, 2002). Deutsch and Hayden have shown that non-locality can be regarded as *locally inaccessible information* and such conception can clarify the decoherence processes (Deutsch and Hayden, 2000; Hewitt-Horsman & Vedral, 2007). In Bohm-Hiley Theory - which has to be kept separate from the minimalist and mechanic reading often made (Berndl et al., 1995)- the radical not-classical aspects are likewise considered as *a kind of information* which is *unknown in the classical conception* and linked to the fundamentally not-local structure of the quantum world. The irreducibility of **R** is thus the consequence of the introduction of quantities depending on the quantum potential and representing not casual events, but rather "potentialities" in a formally very strict sense.

The class (C) instead includes all those theories which tend to reconcile **U** with **R** by introducing new physical process, the so-called **OR** (*Objective Reduction*) (see for ex. Longtin & Mattuck, 1984; Kàrolyhàzy 1974; 1986; Pearle, 1989; Ghirardi et al., 1986; Diòsi, 1989; Percival, 1995; Penrose, 1996a). The **OR** theories focus on the relationship between quantum objects and classical, macroscopic measurement apparatuses (i.e. made up of a number of particles equal to at least $10^{23}$ ). The structure of **U** is modified by introducing new terms and parameters able to obtain a *spontaneous localization*. The current state of these theories is very complex. In these approaches some conflicts with Relativity and suggestions for the quantum gravity stay side by side, so making difficult to say what is *ad hoc* and what will turn out to be fecund. Nevertheless, an acknowledgement is due to the theories of class ( C ) because they identify the asymmetry between **U** and **R** as a radical emergence within the structure of physics and it is a demand for new theoretical proposals able to comprehend the border between quantum and classical systems. However, we have to say that the theories of spontaneous localization do not introduce any new element useful to understand the non-local correlations, but they make the question much more intricate. In fact, if we consider an EPR-Bell experiment with quantum erasing, it is difficult to reconcile it with the thermodynamic irreversibility implicit in the localization concept (see Yoon-Ho et al., 2000; and also Callaghan & Hiley, 2006a ; 2006b).

Obviously, the standpoints on the singular role of **R** in the axioms of QM are much more elaborate, and there are also theories with intermediate features with respect to the abovementioned three classes. In the Cini-Serva theory of correspondence between QM and classical statistical mechanics, for instance, the measurement effect is just to reduce the statistical ensemble and consequently our uncertainty on the system's conditions as well. It is thus eliminated a physically meaningless superposition so as to find the minimum value of the wave packet compatible with quantum indeterminacy (Cini & Serva, 1990; 1992). This theory comprises some typical features of both the class (B) "realism" – yet rejecting its analysis about the nature of the "hidden" information - and the class ( C ), since it is centered on the localization problem at the edge between micro and macro physics, but it does not introduce any new reduction mechanism. Finally – like in the standard interpretation of the class (A) - , in the Cini-Serva theory, the not-separability aspects of quantum correlations are regarded not as a background dynamics, but as being due to the intrinsical casual character of quantum behaviors. The *naïve* (i.e. pre-Bell) theory of hidden variables can be considered as the precursor of the class (B) philosophy, even if the idea of restoring the classical mechanical features has been replaced today by a *not-mechanical* conception in Bohm sense (Bohm, 1951).

*4.3) Quantum Observers and Non-local Information*

Let's try now to focus on the formal facets of QM with respect to the observer. In the quantum context it is impossible for the observer a biunivocal correspondence between the notion of system state, described by the state vector, and the value assignment to the state variables at any instant; it imposes to give up the *indifference principle* of the system state with respect to the observer, typical of the classical systems, and it leads to the breakdown of the causal, local determinism. In QM formalism, such aspect is expressed by the fact that a closure relation is only valid for the eigenvalues (see for ex. Heylighen, 1990).So we can naturally characterize the *quantum systems as informational open systems with respect to the observer.* Now the problem is what meaning we have to give to such logical openness. To be more precise, if we want both no modification in the formalism and a broader comprehension of the QM non-local features, we have to focus our attention on the (B)-type theories and to define a suitable *non-local and classically irreducible information* able at the same to extend the concept of intrinsic computation to quantum systems too. Once again, computation theory shows to be very useful in characterizing the physical systems. In fact, considering the irreducibility of an outcome **R** to the evolutionary structure of **U**, we can state that it is impossible to apply an algorithmic causal structure to two quantum events relative to an observer O at different times. We can so conclude that *a quantum system is a system where the quantum events cannot be correlated one with the other by means of a Shannon-Turing-type computational model.* This proves the intrinsically "classic" roots of the computation concept. We will see in the following that the Bohm-Hiley active information is a really useful approach. But first we have to briefly review the emergent processes in Quantum Field Theory (QFT).

5. **Emergence in QFT**

The idea according to which the QFT distinctive processes are those exhibiting intrinsic emergence and not mere detection of patterns is widely accepted by the community of physicists by now (Anderson and Stein, 1985; Umezawa, 1993; Pessa, 2006; Vitiello, 2001; 2002). The central idea is that in quantum systems with infinite states, as it is known, different not unitarily equivalent representations of the same system, and thus phase transitions structurally modifying the system, are possible. This takes place by means of *spontaneous symmetry breaking* (SSB), i.e. the process which changes all the fundamental states compatible with a given energy value. Generally, when a variation of a suitable parameter occurs, the system will switch to one of the possible fundamental states, so breaking the symmetry. This causes a balance manifesting itself as long-range correlations associated with the Goldstone-Higgs bosons which stabilize the new configuration. The states of bosonic condensation are, in every respect, forms of the system's *macroscopic coherence*, and they are peculiar of the quantum statistics, formally depending on the *indistinguishability of states* with respect to the observer. The new phase of the system requires a new descriptive level to give account for its behaviors, and we can so speak of intrinsic emergence. Many behaviors of great physical interest such as the phonons in crystal, the Cooper pairs, the Higgs mechanism and the multiple vacuum states, the inflation and the "cosmic landscape" formation in quantum cosmology can be included within the SSB processes. It is so reasonable to suppose that the fundamental processes for the formation of structures essentially and crucially depend on SSB.

On the other hand, the formal model of SSB appears problematic in many respects when we try to apply it to systems with a finite number of freedom degrees. Considering the neural networks as an approximate model of QFT is an interesting approach (Pessa & Vitiello, 1999; 2004).The generalized use of the QFT formalism as a theory of emergent processes requires new hypotheses about the system/environment interface. The most ambitious challenge for this formalism is surely the *Quantum Brain* Theory (Ricciardi & Umezawa, 1967; Vitiello, 2001).

A formally interesting aspect of QFT is that it allows, within limits, to frame the question of reductionism in a clear and not banal way, i.e. to give a clear significance to the relations among different descriptive levels. In fact, if we define the phenomenology linked to a given range of energy and masses as the descriptive level, it will be possible to use the renormalization group (RG) as a *tool of resolution* to pass from a level to another by means of a variation of the group's parameters. We thus obtain a succession of descriptive levels - a tower of *Effective Field Theory (EFT)* - having a fixed *cut-off* each, able to grasp the peculiar aspects of the investigated level (Lesne, 1998; Cao, 1997; Cao & Schweber, 1993). In this way each level is connected to the other ones by a rescaling of the kind $\Lambda_0 \to \Lambda(\sigma) = \sigma\Lambda_0$, where $\Lambda_0$ is the cut-off parameter relative to the fixed scale of the energies/masses into play. The universality of the SSB mechanism is deeply linked to such aspect, and the possibility to use the QFT formalism as a general theory of emergence puts the problem of extending the EFT "matryoshka-like" structures allowed by the RG to different systems.

In what sense can the intrinsic emergence of SSB be compared to the phenomenological detection of patterns and which are the radically quantum features in the same sense we pointed out for ordinary QM? Also in the case of the SSB processes, the phase transition is led by an order parameter towards a globally predictable state, i.e. we know that there exists a value beyond which the system will reach a new state and will exhibit macroscopic correlations. Once again a prominent role is played by the boundary conditions (all in all, a phonon is the dynamic emergence occurring within a crystal lattice, but it is meaningless out of the lattice). Moreover, in the SSB there exists a transient phase whose description is widely classic. Where the analogy falls down and we can really speak of an irreducibly not-classic feature is in bosonic condensation which is a non-local phenomenon. While in a classical dissipative system is possible, in principle, to obtain information about the "fine details" of a bifurcation and to know where "the ball will fall", in a process of SSB it is impossible for reasons connected to the nature of the *quantum roulette*! In this sense the *radical features* of emergence in QFT and in QM are of the same nature and they demand a new information theory able to take into account the non-local aspects.

## 6. Quantum Information from the Structure of Quantum Phase Spaces

The concept of active information has been developed by Bohm and Hiley and the Birbeck College group within a wide research programme aimed at comprehending the QM non-local features (see for ex. Bohm & Hiley, 1995; Hiley, 1991; Hiley et al., 2002; Monk & Hiley, 1993; 1998; Brown & Hiley, 2004). This theory is formally equivalent to the standard one and does not introduce any additional hypothesis. The features often described as "Bohmian mechanics" and aimed at a "classic" visualization of the trajectories of quantum objects are not essential. The theory rather tends to grasp the essentially *not-mechanic* nature of quantum processes (Bohm, 1951), in the sense of a topology of not-separability quite close, in formalism and spirit, to non-commutative geometries (Demeret et al., 1997; Bigatti, 1998). It is known that the theory first started from splitting the Schrödinger ordinary equation in real and imaginary part in order to obtain the quantum potential $Q(r,t) = -(\hbar^2/2m)(\nabla^2 R(r,t)/R(r,t))$. This potential QP, unknown in classical physics, contains *in nuce* the QM non-local features and individuates an infinite set of phase paths; in particular, the QP gets a contextual nature, that is to say it brings global information about the quantum system and its environment. We underline that such notion is absolutely general and can be naturally connected to the Feynman path integrals. Our aim is to briefly delineate the geometrical aspects of the quantum phase spaces from which the QP draws its physical significance.

Let us consider an operator $O$ and the wave function $\psi(O)$. By using the polar form of the wave function we can write the usual equations of probability and energy conservation in operatorial form:

$$(1) \qquad i\frac{d\rho}{dt} + [\rho, H]_- = 0; \quad \rho\frac{dS}{dt} + \frac{1}{2}[\rho, H]_+ = 0,$$

where $\rho = \Psi^*(O)\rangle\langle\Psi(O)$ is the density operator which makes available the entropy S of the system as $S = tr\rho\ln\rho$, i.e. as the maximum obtainable information from the system by means of a complete set of observations (Aharonov & Anandan, 1998). Now let us choose a *x*-representation for the (1) and we will get:

(2) $$\frac{\partial \rho}{\partial t}+\nabla_r j=0\,;\,\frac{\partial S}{\partial t}+\frac{(\nabla_r S)^2}{2m}+Q(r,t)+V(r,t)=0.$$

Instead, in a *p*-representation the (1) take the form:

(3) $$\frac{\partial \rho}{\partial t}+\nabla_p j=0\,;\,\frac{\partial S}{\partial t}+\frac{p^2}{2m}+Q(p,t)+V(\nabla_p S,t)=0.$$

From the probability conservation in (2) and (3), two sets of trajectories in the phase space can be finally derived, with both R and Re for real:

(4) $$\nabla_r S=\text{Re}\bigl[\Psi^*(r,t)P\Psi(r,t)\bigr]=p_R\,;\,\nabla_p S=\text{Re}\bigl[\Phi^*(p,t)X\Phi(p,t)\bigr]=x_R.$$

The conceptual fundamental point is that we obtain the quantum potential *only when we have chosen a representation* for the (1). Such a construction is made necessary by the non-commutative structure of the phase spaces of the conjugate variables, and it implies that the observer has to follow a precise process of *information extracting* via the procedure *state preparation→state selection→measurement*. So the quantum potential can be regarded as the measure of the *active information* extracted from what Bohm and Hiley have called *implicate order* to the information *prepared* for the observation in the *explicate order*. The *process algebra* (Symplectic Clifford algebra) between implicate and explicate order is thus a *dynamics of quantum information*.

It is worth noticing that in the (4), given a representation, the conjugate variable is a "beable" variable (Bell, 1987), i.e. a construction depending on the choice made by the observer. Such choice is in no way "subjective", but it is deeply connected with the phase space structure in QM. The Bohm and Hiley's reading restores the natural symplectic symmetry between *x* and *p*, but, unlike the classical case, it does so by geometrically justifying the *complementarity* notion.

To better understand the dynamics of quantum information and its relation with Shannon-Turing classical information, let us consider the notions of both Implicate and Explicate Order and enfolded and unfolded information. (Licata & Morikawa, 2007).

For our aims, it will suffice here to define the Implicate Order as the non-commutative structure of the conjugate variables of the quantum phase space. Therefore, it is impossible to express such structure in terms of space-time without making a choice within the phase space beforehand, so fixing an explicate order. Let us designate the implicate order with E and the explicate order with E'; we will say that E is a source of enfolded information, whereas E' contains the unfolded information extracted from E. The relation between the two orders is given by:

(5) $$E'(t)=\int_{implicate} G(t,\tau)E(\tau)d\tau,$$

where $G(t,\tau)$ is a Green's Function and $\tau$ is the unfolding parameter. The inverse operation, from the explicit space-time structure to the implicate order, is so given by:

(6) $$E(\tau)=\int_{explicate} G(\tau,t)E'(t)dt$$

The passage from (5) to (6) – and vice versa - has a simple physical significance. The unfolding corresponds to the *state selection*, and the enfolding to the *state preparation.* Let us consider a typical two-state system, for ex. a spin ½ particle:

$$(7) \qquad |\Psi\rangle = a|\uparrow\rangle + b|\downarrow\rangle.$$

Here *a* and *b* are a measure of the active information extracted from the implicate order by choosing the state variable. It has to be noticed that the state preparation itself , as well as the measurement, modifies the system contextual information, so defining a new relationship between the background order $E(\tau)$ and the foreground one $E'(t)$. In other words, being chosen a variable, the information of the previous explicate order vanishes into the implicate order.

During the measurement, information becomes enactive, that is to say that the information contained in the superposition state (7) is destroyed, thus selecting between *a* and *b* via an artificial unfolding. Here the term "artificial" means that the measurement - or the "collapse" – has not a privileged position within the theory, but it is only one of the ways which the configuration of the system active information can change into. For example, a creation and annihilation process of particles can be considered as a "spontaneous" process of unfolding/enfolding. What is really important is the *relationship between the contextuality of active information and the non-commutativity of the phase space*. We can sum up all this by saying that in Bohm-Hiley theory a *quantum event* is the expression of a deeper *quantum process* connecting the description in terms of space and time with the intrinsic non-local one of QM. We can say that the implicate order is a realization of the Wheeler "pre-space" (Wheeler, 1980).

The quantum potential, as well as the appearing of "trajectories" in the (4), in no way restores the classic view. In fact, the algebraic structure of QM clearly tells us that *both* the sets of trajectories are necessary to comprehend the quantum processes, and describing them in the explicate order implies a complementarity, and consequently a structural loss of information. Any pretence about the centrality of *x*-representation is arbitrary and it is only based on a classical prejudice , i.e. the "position" regarded as the "existence". In no way a quantum system has to be considered as less "real" and "objective" than a classical one. It is the observer role which changes just in relation to the peculiar nature of quantum processes. So the trajectories have not to be regarded as "mysterious" or "surreal" lines of force violating Relativity, but as informational configurations showing the intrinsic not-separability of the quantum world in the space-time foreground.

From what we have said above, it clearly appears that the problem of the emergence of the classical world from the quantum one cannot be dealt with by using a classical limit for the quantum potential, despite the statistical interest of such pragmatic approach (see for ex. Allori & Zanghì, 2001). The limit of classicality has rather to be faced as a problem of relationships between algebra and metric in the sense of the principle of reconstruction of Gel'fand (Demaret et al., 1997). Another significant point, which we can only mention here, is the importance of the unfolding parameter in quantum cosmology, for example in passing from a timeless quantum De Sitter-like universe to a post inflationary time evolution (Callander & Weingard, 1996; Nelson Pinto-Neto, 2000; Lemos & Monerat, 2002; Licata, 2006; Chiatti & Licata, 2007).

The Shannon-Turing information comes into play when a syntactic, local analysis of the system is possible on a well-defined channel. In QM, it is possible only when a system has been prepared in a set of orthogonal wave functions, thus making a selection of the enfolded information which can be analyzed in terms of usual quantum bits. Elsewhere we have pointed out how such crucial difference between contextual active information and q-bit information can be used to explain forms of quantum computation much more powerful than the one based on quantum gates (Licata, 2007).

A different and more immediate way to understand the limits of the classical computation theory in QM is to take into consideration again the trajectories in the (4). As the two sets are complementary, the thought experiment of "rewinding" the trajectories of an unmeasured quantum system is structurally unable to provide information about the details of evolution. In fact, let us suppose we have chosen, for example, a family of trajectories in *x*-representation and we – yet ideally – have made a computational analysis on it, all the same we can say nothing on the computation of the trajectories in the *p*-representation: just like the variable on which it is centred on, the computation on *p*-representation appears to be a "beable computation"! It brings out a deep connection between non-commutativity, non-locality and the new kind of quantum time-asymmetry related to the fact that the logic of the process of unfolding/enfolding described by (5) and (6) – that is to say $\tau \to t$ and vice versa- implies an uncomputable modification of the system information.

We can conclude that the features of informational openness of QM are a direct consequence of the non-commutative algebra which imposes upon the observer to make a choice on the available information and it prevents from describing the intrinsic computation of a quantum systems via a Shannon-Turing model because of the contextual nature of the active information.

## 7. On the relationships between physics and computation

In the last years an important debate on the relationships between physical systems, formal models and computation has been developed. In our analysis, we have used the classical computation theory not only as simulation means, but rather as a conceptual tool able to let us understand the formal relationships between the system's behaviours and the information that the observer can get about it. We have seen that it is possible to give a computational description of the classical systems centred on the possibility to be always able to identify the information in local way. So, emergence in classical systems is fundamentally of computational kind and there exists a strict correspondence between classical systems and computational dynamics (Baas & Emmeche, 1997). This is not in contradiction with the appearing of non-Turing features in some classical systems (see for ex. Siegelmann, 1998; Calude, 2004) because it has been proved that sets of interactive Turing Machines (TMs) with oracles can show computational abilities superior than those of a TM in the strict sense (Kieu & Ord, 2005; Collins, 2001). Actually, if we look at the "oracles" as metastable configurations fixed during the evolution of a system, this is the minimum required baggage to understand the biological evolution and mind. QM *does not seem to be necessary* to comprehend the essential features of life and cognition, in contrast to what Penrose claims (Penrose, 1996).

In QM, we deal instead with radical processes of observational emergence which cannot be overcome by the construction of a new model. The Bohm-Hiley theory helps us to understand such radical features of QM by the Implicate/Explicate Order process algebra and the non-commutative structure of the quantum phase spaces. The failure of a TM-based observer in describing quantum systems and the consequent recourse to a probabilistic structure are related to the fact that any observer belongs to the explicate order and it has to use a space-time structure to build up the concept of physical event. From this view, the role of computation in physics appears to be even more profound than what the Turing Principle suggests (Deutsch, 1998). In fact, *the Shannon-Turing theory here appears as a computability general constraint on the causal structures which can be defined in space-time* and as the extreme limit of any description in terms of space-time (Markopoulou-Kalamera, 2000a; 2000b; Tegmark, 2007).

Processing information is what all physical systems do. The way to process information depends on the system's nature, and it is not surprising that the Shannon-Turing information naturally fits to not-quantum systems or, at the utmost, to q-bit systems because it has been conceived within a classical context. In quantum systems instead, the breakdown of the classical computation model calls into play a new concept of information, in the same way as more than once in the history of physics, a conceptual crisis has required new strategies to comprehend the unity of the physical world. The Bohm-Hiley concept of active information allows to extend the concept of intrinsic computation to quantum systems, too. In order to make such notion efficacious, a great deal of new ideas will be necessary. Already in 1935, Von Neumann wrote in a letter to Birkhoff: "I would like to make a confession which may seem immoral: I do not believe in Hilbert space anymore." (Von Neumann, 1935). In fact the separable Hilbert spaces are not fit for representing infinite systems far from equilibrium. It is thus necessary a QFT able, at least, to represent the semigroups of irreversible transformations – i.e. operators for the absorption and dissipation of quanta – by not-separable Hilbert spaces. Here, a new scenario connecting non-commutativity, irreversibility and information emerges.

The "virtuous circle" of systems-models-computation finds its ultimate significance not only in the banal fact that we build models of the physical world so as to be "manageable", but also because we are constrained to built such models based on family of observers in the explicate order. It suggests that an authentic "pacific co-existence" between Relativity and QM will require a general mechanism to obtain space-time transformation groups via an unfolding parameter as a boost from a more complex underlying background structure. In this context, the hyperspherical group approach appears to be interesting (see Licata, 2006; Chiatti & Licata, 2007).

The physical world has no need for observers for its structure and evolution. But the nature of the quantum processes makes the relationship between the observer and the observed irreducibly *participatory*, such that the description of any physical system is necessarily influenced by unbounded and contextual information. Such kind of information is not cooped up in the "Hilbert cage" and has less to do with what is usually meant by quantum computing. Far from being a "shadow" information, it fixes the fundamental non-local character of the quantum world and the limit of Shannon-Turing information in describing the intrinsic computation of quantum systems. In contrast, it is just this radical uncomputable feature which individuates the Boolean characteristics of the observers in the space-time and allows, thanks to the spontaneous symmetry breaking processes, the increasing complexity of the physical universe.